\documentstyle [12pt] {article}

\catcode`@=12
\begin {document}
\thispagestyle {empty}
\begin{flushright} UCRHEP-T105\\February 1993\
\end{flushright}
\vspace{0.5in}
\begin{center}
{\Large \bf Derivation of $m_A \simeq M_Z$ and $\tan \beta > \sqrt 3$ in the\\
Minimal Supersymmetric Standard Model\\}
\vspace{1.5in}
{\bf Ernest Ma\\}
\vspace{0.3in}
{\sl Department of Physics\\}
{\sl University of California\\}
{\sl Riverside, California 92521\\}
\vspace{1.5in}
\end{center}
\begin{abstract}\
In the minimal supersymmetric standard model, the Higgs sector has two
unknown parameters, usually taken to be $\tan \beta \equiv v_2/v_1$ and
$m_A$, the mass of its one physical pseudoscalar particle.  By minimizing
the minimum of the Higgs potential along a certain direction in parameter
space, it is shown that $m_A = M_Z$ + radiative correction, and if one
further plausible assumption is made, $\tan \beta > \sqrt 3$.
\end{abstract}

\newpage
\baselineskip 24pt

If there is physics beyond the standard ${\rm SU(2) \times U(1)}$ electroweak
gauge model, supersymmetry is generally considered to be an excellent
candidate.  The minimal supersymmetric standard model\cite{1} must then
have two scalar doublets  $\Phi_1 = (\phi_1^+, \phi_1^0)$ and $\Phi_2 =
(\phi_2^+, \phi_2^0)$, with Yukawa interactions $(\overline {u,d})_L d_R
\Phi_1$ and $(\overline {u,d})_L u_R \tilde \Phi_2$ respectively, where
$\tilde \Phi_2 = i \sigma_2 \Phi_2^* = (\overline {\phi_2^0}, -\phi_2^-)$.
Hence the mass matrix for quarks of charge $-1/3$ ({\it i.e.} $d$, $s$, and
$b$) is proportional to the vacuum expectation value $\langle \phi_1^0
\rangle \equiv v_1$, and that for quarks of charge 2/3 ({\it i.e.} $u$, $c$,
and $t$) is proportional to $\langle \phi_2^0 \rangle \equiv
v_2$.  There are two unknown parameters in the Higgs sector of this model.
They are usually taken to be $\tan \beta \equiv v_2/v_1$ and $m_A$, the mass
of its one physical pseudoscalar particle.  Numerous phenomenological
studies\cite{2} have been made using these as continuous variables.  However,
there is a plausible theoretical argument for $m_A = M_Z$ at tree level and
perhaps also $\tan \beta > \sqrt 3$, as shown below.

The Higgs potential of the minimal supersymmetric standard model is given
by\cite{3}
\begin{eqnarray}
V &=& m_1^2 \Phi_1^\dagger \Phi_1 + m_2^2 \Phi_2^\dagger \Phi_2 - m_{12}^2
(\Phi_1^\dagger \Phi_2 + \Phi_2^\dagger \Phi_1) \nonumber \\ &+& {1 \over 8}
(g_1^2 + g_2^2) \left[ (\Phi_1^\dagger \Phi_1)^2 + (\Phi_2^\dagger \Phi_2)^2
\right] \nonumber \\ &-& {1 \over 4} (g_1^2 - g_2^2) (\Phi_1^\dagger \Phi_1)
(\Phi_2^\dagger \Phi_2) - {1 \over 2} g_2^2 (\Phi_1^\dagger \Phi_2)
(\Phi_2^\dagger \Phi_1),
\end{eqnarray}
where $g_1$ and $g_2$ are the U(1) and SU(2) gauge couplings respectively.
Let $V$ be broken spontaneously, then
\begin{equation}
m_1^2 - m_{12}^2 v_2/v_1 + {1 \over 4} (g_1^2 + g_2^2) (v_1^2 - v_2^2) = 0,
\end{equation}
and
\begin{equation}
m_2^2 - m_{12}^2 v_1/v_2 + {1 \over 4} (g_1^2 + g_2^2) (v_2^2 - v_1^2) = 0.
\end{equation}
Hence
\begin{equation}
m_1^2 + m_2^2 = m_{12}^2 (\tan \beta + \cot \beta),
\end{equation}
and
\begin{equation}
v_1^2 + v_2^2 = {{-4 m_1^2 \cos^2 \beta + 4 m_2^2 \sin^2 \beta} \over
{(g_1^2 + g_2^2) (\cos^2 \beta - \sin^2 \beta)}}.
\end{equation}
There are five physical scalar particles with masses given by
\begin{equation}
m_A^2 = m_1^2 + m_2^2,
\end{equation}
\begin{equation}
m_{H^\pm}^2 = M_W^2 + m_A^2,
\end{equation}
and
\begin{equation}
m_{H_1^0,H_2^0}^2 = {1 \over 2} \left[ M_Z^2 + m_A^2 \pm \sqrt {(M_Z^2 +
m_A^2)^2 - 4 M_Z^2 m_A^2 \cos^2 2 \beta} \right],
\end{equation}
where
\begin{equation}
M_W^2 = {1 \over 2} g_2^2 (v_1^2 + v_2^2)
\end{equation}
and
\begin{equation}
M_Z^2 = {1 \over 2} (g_1^2 + g_2^2) (v_1^2 + v_2^2)
\end{equation}
are the squares of the masses of the $W$ and $Z$ bosons.

The part of $V$ involving only neutral fields depends on four parameters:
$m_1^2$, $m_2^2$, $m_{12}^2$, and $g_1^2 + g_2^2$.  At its minimum $V_0$,
we can choose to keep $m_1^2$ and $m_2^2$, but replace $m_{12}^2$ by
$\tan \beta$ through Eq. (4) and $g_1^2 + g_2^2$ by $v_1^2 + v_2^2$ through
Eq. (5).  In that case,
\begin{equation}
V_0 = {1 \over 2} (v_1^2 + v_2^2) (\cos^2 \beta - \sin^2 \beta) (m_1^2 \cos^2
\beta - m_2^2 \sin^2 \beta).
\end{equation}
We now seek to minimize $V_0$ in parameter space. This is a reasonable
procedure in that whatever dynamical mechanism is responsible for the soft
breaking of supersymmetry, it may well be such that the lowest possible
value of $V_0$ is automatically chosen.  It is also clear that $v_1^2 +
v_2^2$, $m_1^2$, and $m_2^2$ set the energy scale of the symmetry breaking
and $V_0$ has no lower bound as a function of these parameters.  We should
therefore consider them as fixed and vary $\sin^2 \beta$ to minimize $V_0$.
Let $x \equiv \sin^2 \beta$, then
\begin{equation}
{{\partial V_0} \over {\partial x}} = {1 \over 2} (v_1^2 + v_2^2) \left[
-(3m_1^2 + m_2^2) + 4(m_1^2 + m_2^2) x \right],
\end{equation}
and
\begin{equation}
{{\partial^2 V_0} \over {\partial x^2}} = 2 (v_1^2 + v_2^2) (m_1^2 + m_2^2).
\end{equation}
Hence the minimization of $V_0$ is achieved if
\begin{equation}
x = {{3m_1^2 + m_2^2} \over {4(m_1^2 + m_2^2)}}
\end{equation}
and $m_1^2 + m_2^2 > 0$ which is consistent with Eq. (6).  Using Eq. (5), we
then find
\begin{equation}
m_1^2 + m_2^2 = {1 \over 2} (g_1^2 + g_2^2) (v_1^2 + v_2^2),
\end{equation}
or equivalently
\begin{equation}
m_A = M_Z.
\end{equation}
This implies
\begin{equation}
m_{H^\pm} = (M_W^2 + M_Z^2)^{1 \over 2} \simeq 121~{\rm GeV}
\end{equation}
and
\begin{equation}
m_{H_1^0,H_2^0} = M_Z (1 \pm \sin 2 \beta)^{1 \over 2}.
\end{equation}

{}From Eq. (14), we find
\begin{equation}
\tan^2 \beta = {{3 m_1^2 + m_2^2} \over {m_1^2 + 3 m_2^2}},
\end{equation}
which shows that if $m_1^2 > 0$ and $m_2^2 > 0$, then $1/\sqrt 3 < \tan
\beta < \sqrt 3$.  However, because $\Phi_2$ couples to the $t$ quark with
a large Yukawa coupling, $m_2^2$ is expected to differ from $m_1^2$ by a
large negative contribution from the renormalization-group equations,\cite{4}
hence the case $m_1^2 > 0$ and $m_2^2 < 0$ should be considered.  We then
obtain
\begin{equation}
\tan \beta > \sqrt 3,
\end{equation}
where $m_1^2 > 3 |m_2^2|$ has also been assumed or else $V_0$ would have been
minimized at $\sin^2 \beta > 1$ which is impossible.  Using Eq. (18), we
find $m_{H_1^0} > 33$ GeV and $m_{H_2^0} < 125$ GeV, with the constraint that
$m_{H_1^0}^2 + m_{H_2^0}^2 = 2 M_Z^2$.  Experimentally, there is no evidence
for the existence of any of the five scalar particles of the minimal
supersymmetric standard model from $Z$ decay or in any other process.
Detailed phenomenological analyses\cite{2} have concluded that for $m_A =
M_Z$, $\tan \beta > \sqrt 3$ is allowed.  The only possible exception
is from the experimental upper bound on the process $b \rightarrow s \gamma$,
which excludes $m_A$ up to 130 GeV according to one calculation.\cite{5}
However, another calculation\cite{6} gives a value which is roughly 20\%
lower, in which case $m_A = M_Z$ and $\tan \beta > \sqrt 3$ are again
allowed for $m_t \leq 150$ GeV.

Using Eq. (5), we could also have written Eq. (11) in the more familiar form
\begin{equation}
V_0 = - {1 \over 8} (g_1^2 + g_2^2) (v_1^2 + v_2^2)^2 \cos^2 2 \beta.
\end{equation}
This means that if we were to fix $g_1^2 + g_2^2$ and $v_1^2 + v_2^2$, the
minimization of $V_0$ would occur at either $\sin \beta = 0$ ($v_2 = 0$)
or $\cos \beta = 0$ ($v_1 = 0$).  Note that because of Eq. (5), we could
not have also kept both $m_1^2$ and $m_2^2$ fixed.  Whereas this solution
does indeed correspond to a local minimum along a particular direction in
parameter space, it is clearly not acceptable phenomenologically.  It also
requires $m_{12}^2$ to be zero which is perhaps not so acceptable
theoretically.  As it is, we have discovered another direction (and
arguably the only one) along which $V_0$ is minimized with both $v_1$ and
$v_2$ nonzero.

If the Yukawa coupling of the $t$ quark to $\Phi_2$ is large, there is
also a significant radiative contribution to the $(\Phi_2^\dagger \Phi_2)^2$
term in the Higgs potential $V$ of Eq. (1).  This generates an extra term in
Eq. (3).  In that case, Eq. (7) remains valid but Eq. (8) is modified so that
in principle $m_{H_2^0} > M_Z$ would be possible.\cite{7}  Here we find
\begin{equation}
m_A^2 \simeq M_Z^2 (1 + \delta_1),
\end{equation}
and
\begin{equation}
m_{H_1^0,H_2^0}^2 \simeq M_Z^2 \left(1 + \delta_2 \pm \sqrt {(1 + \delta_1)
\sin^2 2 \beta + \delta_2^2} \right),
\end{equation}
where
\begin{equation}
\delta_1 = \left( {{3-2\sin^2 \beta} \over {2\sin^2 \beta -1}} \right) \delta,
{}~~~~\delta_2 = \delta + {1 \over 2} \delta_1,
\end{equation}
and
\begin{equation}
\delta = {{3 g_2^2 m_t^4} \over {16 \pi^2 M_W^2 M_Z^2 \sin^2 \beta}}
\ln {\tilde M^2 \over m_t^2},
\end{equation}
$\tilde M$ being an effective mass for the scalar supersymmetric partners of
the $t$ quark and $\tilde M^2 >> m_t^2$ has been assumed.

In conclusion, it has been shown in this note that $m_A = M_Z$ (+ radiative
correction) and $\tan \beta > \sqrt 3$ are plausible results in the minimal
supersymmetric standard model. The key theoretical assumption is that the
minimum $V_0$ of the Higgs potential $V$ should be also a local minimum in
parameter space along a certain direction.  Specifically, if we fix $m_1^2$,
$m_2^2$, and $v_1^2 + v_2^2$, then $V_0$ is minimized at tree level with
$\tan^2 \beta = (3m_1^2 + m_2^2)/(m_1^2 + 3m_2^2)$, resulting in $m_A = M_Z$.
However, radiative correction due to a large value of $m_t$ changes that to
$m_A \simeq M_Z (1 + \delta_1)^{1/2}$ as given by Eq. (22).  If we assume
further that $m_1^2 > 0$ but $m_2^2 < 0$ because of a large negative
contribution from the renormalization-group equations, then $\tan \beta >
\sqrt 3$ is also obtained.
\vspace{0.3in}
\begin{center} {ACKNOWLEDGEMENT}
\end{center}

The author acknowledges the indispensable help of V. Barger in clarifying
the current phenomenological situation with regard to $\tan \beta$ and
$m_A$.  This work was supported in part by the U. S. Department of Energy
under Contract No. DE-AT03-87ER40327.

\newpage
\bibliographystyle{unsrt}

\end {document}